\documentclass[10pt]{article} 
\usepackage{amsmath,amsfonts}
\usepackage{algorithmic}
\usepackage{algorithm}
\usepackage{array}
\usepackage[caption=false,font=normalsize,labelfont=sf,textfont=sf]{subfig}
\usepackage{textcomp}
\usepackage{stfloats}
\usepackage{url}
\usepackage{verbatim}
\usepackage{graphicx}
\usepackage{cite}
\usepackage[colorlinks=true,linkcolor=black,anchorcolor=black,citecolor=black,filecolor=black,menucolor=black,runcolor=black,urlcolor=black]{hyperref}

\usepackage{booktabs}
\usepackage{multirow}

\usepackage{xcolor}

\usepackage[margin=1in]{geometry}
\vspace{-30 cm}

\begin{document}


\title{Corrections to Friis noise factor formulas for cascade networks}

\author{Ankitha E. Bangera \\
\\
\it{Department of Electrical Engineering,} \\
\it{Indian Institute of Technology Bombay,} \\
\it{Mumbai$-$400076, India} \\
E-mail: ankitha\_bangera@iitb.ac.in; ankitha.bangera@iitb.ac.in}
\date{}

\maketitle
\thispagestyle{empty}

\begin{abstract}
The signal-to-noise ratio of a multistage cascade network is often estimated using the well-known Friis' formulas for noise factors (or the noise figures in decibels). However, this article addresses the major errors in Friis' noise factor formulas for higher stages. Additionally, we re-derive the correct formulas to calculate the stage-wise noise factors for cascade networks from the basic definition of noise factors. We then present a comparison of our derived formulas with Friis' noise factor formulas. Contrary to Friis' formula, we define the total noise factor of an \textit{n}-stage cascade network as the product of its stage-wise noise factors. We further validate our derived formulas for a cascade network by correlating them with the expressions for a staircase avalanche photodiode. 
\end{abstract}

\section{Introduction}
\label{sec1}

Cascade mechanism is widely used in various domains such as microelectronics \& photonics \cite{bib1,bib2,bib3,bib4,bib5,bib6}, two-dimensional (opto-) electronics \cite{bib1,bib7,bib8}, telecommunications \& signal processing \cite{bib9,bib10}, integrated circuits \cite{bib11,bib12,bib13,bib14,bib15},  artificial intelligence \cite{bib16,bib17,bib18,bib19}, chemical biology \cite{bib20,bib21}, and so on. Some major applications of networks in cascade include photon detection \cite{bib1}, light detection \& ranging \cite{bib1,bib22}, photovoltage multiplication \cite{bib7}, noise attenuation \cite{bib9}, biosensing \cite{bib15,bib21}, medical imaging \cite{bib23,bib24}, anomaly detection \cite{bib16,bib17}, image enhancement \cite{bib18}, and many more. 

To extract the actual signal component at the output of a network, it is critical to calculate its noise factors. In 1944, Friis defined the noise factor of a network as the ratio of its input signal-to-noise ratio (SNR) to its output SNR \cite{bib25}. SNR is a key factor in determining the overall system performance. There have been efforts to improve the design of systems or individual components to attain high SNR \cite{bib26,bib27,bib28,bib29}. This necessitates the accurate estimation of noise factors as well. To date, Friis' noise factor formulas are often used to calculate the noise factors in multistage cascade networks \cite{bib2,bib3,bib11,bib14,bib15}. However, some of the noise factor formulas proposed by Friis are incorrect and thus need a correction.

This article briefly discusses the existing theory and the well-known Friis noise factor formulas for cascade networks \cite{bib25,bib30,bib31,bib32}. We then re-derive the correct stage-wise noise factor expression and the corresponding total noise factor for cascade networks in terms of their stage-wise noise factors. Our derivation is based on the existing theory for cascade networks, where the output signal of the previous `$(x-1)$-th' stage will be the input to the next `$x$-th' stage. We then compare our newly derived noise factor expressions with Friis' formulas, discussing the corrections required for Friis' noise factors. For validation, we correlate our results for a cascade network with those of a staircase avalanche photodiode (APD). In this article, the terms such as `network,' `amplifier,' `circuit,' `system,' `device,' or any similar term used to indicate a cascade `structure' are interchangeable. However, the term `network' is applied to maintain uniformity throughout the article.

\section{Theory} 
\label{sec2}

\begin{figure}[!t]
\centering
\includegraphics[width=5in]{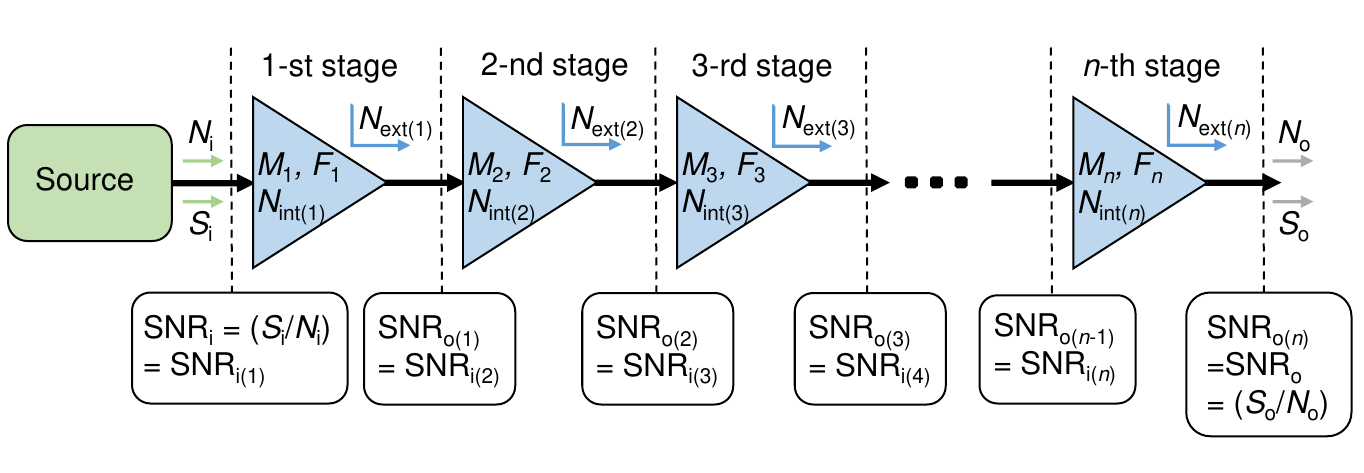}
\caption{Block diagram of an $n$-stage cascade network.}
\label{fig_1}
\end{figure}

Fig.~\ref{fig_1} shows the block diagram of an $n$-stage cascade network. In the block diagram, $S_\text{i}$ is the source's signal power at the input terminals of the cascade network (or input signal power); $N_\text{i}$ is the source's noise power at the cascade network's input terminals (or input noise power); $M_x$ is the gain of the network's $x$-th stage (or stage-wise gains); $G_x=M_x^2$ is the power gain of the network's $x$-th stage (or stage-wise power gains), defined as the ratio of a network's $x$-th stage's output signal power to its input signal power; $F_x$ is the noise factor at the network’s $x$-th stage (or stage-wise noise factors); $N_{\text{int}(x)}$ is the noise power due to the internal noises or irregularities generated within the network's $x$-th stage (or stage-wise internal noise powers), amplified only at the succeeding stages; $N_{\text{ext}(x)}$ is the noise power of the externally added noise at the output of the network's $x$-th stage, such that the noise is amplified only at the succeeding stages; $\text{SNR}_{\text{i}(x)}$ and $\text{SNR}_{\text{o}(x)}$ are the SNRs at the input and output of the network's $x$-th stage; $\text{SNR}_\text{i}$ is the SNR at the input of the cascade network (or input SNR); $\text{SNR}_\text{o}$ is the SNR at the output of the cascade network (or output SNR); $S_\text{o}$ is the signal power at the cascade network's output terminals (or output signal power); and $N_\text{o}$ is the noise power at the cascade network's output terminals (or output noise power). 

\subsection{Existing noise theory and Friis' noise factors (stage-wise internal noises excluded)} 
\label{subsec2_1}

The noise factor of a two-port network proposed by Friis \cite{bib25} is defined as the ratio of the SNR at the network's input terminals to the SNR at the network's output terminals, which was later standardized by IEEE \cite{bib30}. Here, SNR is the ratio of the signal power to the noise power at a particular set of terminals in a two-port network. The power gain of the two-port network ($G$) is defined as the ratio of its output signal power ($S_\text{o}$) to its input signal power ($S_\text{i}$). Haus' \cite{bib31,bib32} further studies on Friis' formulations, express a two-port network's noise factor as the ratio of the noise power at its output terminals ($N_\text{o}$) to the product of the noise power at its input terminals ($N_\text{i}$) and its power gain ($G$). According to Haus \cite{bib31,bib32}, if $N_\text{ext}$ is the noise power of the externally added noise at the output of the two-port network, such that, the noise power at the network's output terminals is equal to the sum of the amplified input noise power and the externally added noise power at its output terminals; then, the two-port network's noise factor is given by,  

\begin{equation}
\label{eqn_1}
\begin{aligned}
F_{\text{T}_1} = \frac{N_\text{o}}{N_\text{i}G} = 1 + \frac{N_\text{ext}}{N_\text{i}G}
\end{aligned}
\end{equation}

However, equation~\eqref{eqn_1} is for a single-stage two-port network. Friis' extended model for two-port networks in cascade \cite{bib25} formulates the total noise factor of an $n$-stage cascade network shown in Fig.~\ref{fig_1}, but without considering the internal noises or irregularities generated within its stages. Therefore, neglecting the stage-wise internal noise powers ($N_{\text{int}(x)}$), Friis' total noise factor formula for cascade networks is given by, 

\begin{equation}
\label{eqn_2}
\begin{aligned}
F_{\text{T}_n} &= 1+\sum_{x=1}^{n}\left(\frac{N_{\text{ext}(x)}}{N_\text{i}\prod_{y=1}^{x}G_y}\right)
\end{aligned}
\end{equation} 

However, according to Friis \cite{bib25} and the IEEE standards \cite{bib30}, the noise factor formula for a cascade network's $x$-th stage (or stage-wise noise factors) is expressed as,  

\begin{equation}
\label{eqn_3}
F_x^{\text{Friis}} = 1+\frac{N_{\text{ext}(x)}}{N_\text{i}G_x}
\end{equation}

Therefore, the total noise factor for an $n$-stage cascade network in terms of its stage-wise noise factors is written as, 

\begin{equation}
\label{eqn_4}
F_{\text{T}_n}^{\text{Friis}} = F_1^{\text{Friis}}+\sum_{x=2}^{n}\left(\frac{F_x^{\text{Friis}}-1}{\prod_{y=1}^{(x-1)}G_y}\right)
\end{equation}

The derivation of equation~\eqref{eqn_4} is included in Appendix~\ref{secA1}.

\subsection{Our new extended noise factor expressions for networks in cascade (stage-wise internal noises included)} 
\label{subsec2_2} 

For cascade networks, the previous stage's output will be the next stage's input. Therefore, as per the definition of the noise factor \cite{bib25,bib30,bib31,bib32}, a cascade network's stage-wise noise factor must be equal to the ratio of SNR at the input of the network's stage to SNR at the output of its corresponding stage. From the block diagram shown in Fig.~\ref{fig_1}, the stage-wise noise factor at the network's $x$-th stage can be equated to,  

\begin{equation}
\label{eqn_5}
F_x = \frac{\text{SNR}_{\text{i}(x)}}{\text{SNR}_{\text{o}(x)}} = \frac{\text{SNR}_{\text{o}(x-1)}}{\text{SNR}_{\text{o}(x)}}
\end{equation}

Solving equation~\eqref{eqn_5}, a cascade network's stage-wise noise factors may be defined as the ratio of the total noise power at the output of the network's $x$-th stage ($N_{\text{o}(x)}$) to the product of the total noise power at the input of its $x$-th stage ($N_{\text{i}(x)}$) and its corresponding stage-wise power gain ($G_x$), expressed as, 

\begin{equation}
\label{eqn_6}
\begin{aligned}
F_x^{\text{Bang}} = \frac{N_{\text{o}(x)}}{N_{\text{i}(x)}G_x} = 1+\frac{\left(N_{\text{int}(x)}+N_{\text{ext}(x)}\right)}{N_{\text{i}(x)}G_x} \neq\Bigl(F_x^{\text{Friis}}\Bigr)
\end{aligned}
\end{equation}

Here, the noise power at the input terminals of the network's $x$-th stage can be formulated as, 

\begin{equation}
\label{eqn_7}
\begin{aligned}
N_{\text{i}(x)} = N_\text{i}\prod_{j=1}^{(x-1)}G_j+\sum_{k=1}^{(x-1)}\Bigl\{\left(N_{\text{int}(k)}+N_{\text{ext}(k)}\right)\prod_{l=k+1}^{(x-1)}G_l\Bigr\} \neq\Bigl(N_{\text{i}}; \forall x\geq2 \Bigr)
\end{aligned}
\end{equation}

Since the definition of the total noise factor of an $n$-stage cascade network is equal to the ratio of its input SNR to its output SNR \cite{bib25,bib30,bib31,bib32}, the total noise factor of an $n$-stage cascade network shown in Fig.~\ref{fig_1} is given by, 

\begin{equation}
\label{eqn_8}
\begin{aligned} 
F_{\text{T}_n} &= 1+\sum_{x=1}^{n}\left(\frac{\left(N_{\text{int}(x)}+N_{\text{ext}(x)}\right)}{N_\text{i}\prod_{y=1}^{x}G_y}\right) 
\end{aligned}
\end{equation}

Therefore, the new expression to determine the total noise factor of an $n$-stage cascade network in terms of its stage-wise noise factors is given by equation~\eqref{eqn_9}, which is not equal to equation~\eqref{eqn_4}. 

\begin{equation}
\label{eqn_9}
F_{\text{T}_n}^{\text{Bang}} = \prod_{x=1}^{n}F_x^{\text{Bang}}\neq\Bigl(F_{\text{T}_n}^{\text{Friis}}\Bigr)
\end{equation}

The derivation of equation~\eqref{eqn_9} is included in Appendix~\ref{secB1}.

\section{Results and Discussion} 
\label{sec3}

From equation~\eqref{eqn_9}, the total noise factor of an $n$-stage cascade network is defined as the product of its stage-wise noise factors, with equal contributions from all its stages. On the contrary, Friis' total noise factor for cascade networks given by equation~\eqref{eqn_4} heavily depends on the stage-wise noise factor of its 1-st stage. 

Furthermore, the noise factors formulated by Friis \cite{bib25} and later re-expressed by Haus \cite{bib31,bib32}, do not include the stage-wise internal noises that are expected to be generated within a practical network's stage. One such example of these stage-wise internal noises is the staircase APD's irregularities (randomness) in the stepwise impact ionization \cite{bib2,bib3,bib33}. By contrast, our newly derived expressions include these stage-wise internal noises, as detailed in subsection~\ref{subsec2_2}. However, to compare our new expressions with Friis' noise factors, we neglect the term $N_{\text{int}(x)}$ corresponding to the stage-wise internal noise powers in subsection~\ref{subsec3_1}. 

Additionally, the externally added stepwise noises are negligible for solid-state devices such as staircase APDs \cite{bib2,bib3,bib33}, which behave similar to a cascade network. The major source of noise in these devices is dominated by the internal noise generated due to the irregularities in the stepwise impact ionization. Therefore, in subsection~\ref{subsec3_2}, we ignore the noise contribution from the externally added stage-wise noise powers corresponding to the term $N_{\text{ext}(x)}$.

\subsection{Comparison with Friis' noise factor formulas for cascade networks (neglecting internally generated stage-wise noises)} 
\label{subsec3_1}

\begin{figure}[!t]
\centering
\includegraphics[width=5in]{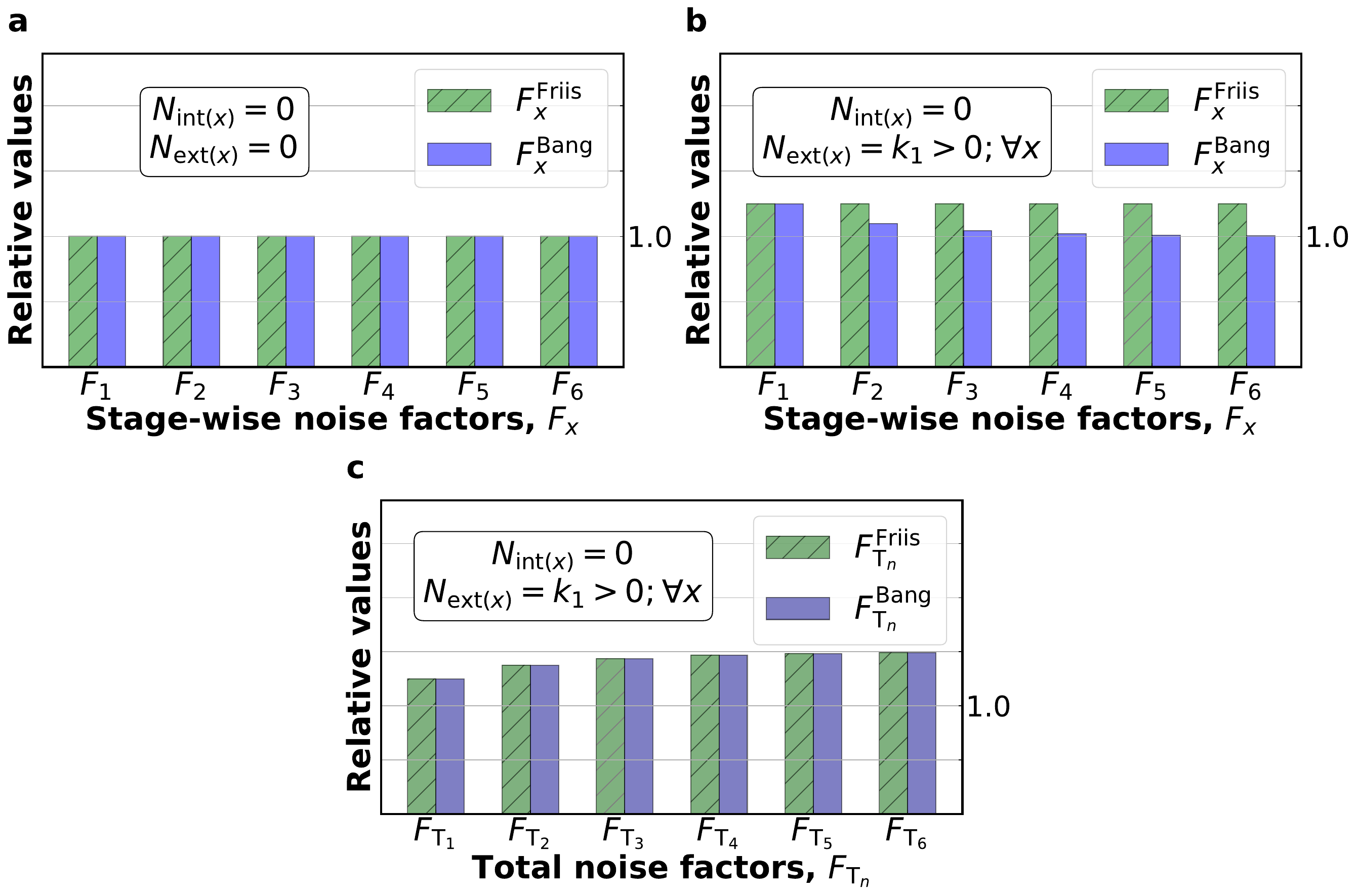}
\caption{A comparison of Friis' and our noise factors for up to the 6-th stage (neglecting internally generated noises). Bar charts comparing Friis' and our stage-wise noise factor estimates (a) if no external noises are added at the outputs of all stages and (b) if there exists a non-zero positive identical external noise added at the output of all stages (assuming $\forall x$, $G_x=G \geq 1$). (c) A bar chart comparing Friis' and our total noise factor estimates for a non-zero positive identical external noise added at the output of all stages (assuming $\forall x$, $G_x=G \geq 1$).} 
\label{fig_2}
\end{figure}

For a single-stage network or a cascade network's 1-st satge $N_{\text{i}(1)}=N_{\text{i}}$, thus, our new expression for stage-wise noise factors given by equation~\eqref{eqn_6} will be equivalent to Friis' equations~\eqref{eqn_3} or~\eqref{eqn_1}. However, for higher stages ($x\ge2$), equation~\eqref{eqn_6} is unequal to equation~\eqref{eqn_3}, except when there are no externally added or internally generated stage-wise noises resulting in unity stage-wise noise factors. This is illustrated in Appendix~\ref{secC1}. A bar chart depicting unity stage-wise noise factors in the absence of internal and external stage-wise noises is shown in Fig.~\ref{fig_2}a. Further, comparing equations~\eqref{eqn_3} and~\eqref{eqn_6}, the major difference lies in the terms $N_{\text{i}}$ and $N_{\text{i}(x)}$. Where $N_{\text{i}}$ is the noise power at the network's input terminals (\textit{i.e.}, the 1-st stage's input terminals of a cascade network). Whereas, $N_{\text{i}(x)}$ given by equation~\eqref{eqn_7} is the noise power at the input terminals of the network's $x$-th stage, which is greater than or equal to the term $N_{\text{i}}$. Therefore, estimates of equation~\eqref{eqn_6} will be less than or equal to those obtained using equation~\eqref{eqn_3}. According to Friis, the stage-wise noise factor values remain the same for all stages, when a non-zero positive identical external noise is added at the output of all stages (assuming $\forall x$, $G_x=G \geq 1$). Whereas, our derived formula predicts that the stage-wise noise factors reduce with the stage number. This reduction is because, as the stage number increases, the total noise at the input of that stage increases due to the external noises added at the outputs of all the previous stages and the input noise that gets amplified at the previous stages. A bar chart comparing the relative values of the stage-wise noise factors for a non-zero positive identical external noise added at the output of all stages (assuming $\forall x$, $G_x=G \geq 1$), calculated using the Friis' formula (equation~\eqref{eqn_3}) and our expression (equation~\eqref{eqn_6}) for up to the 6-th stage is shown in Fig.~\ref{fig_2}b. Thus, for all the higher stages in a cascade network $\left(x\geq2\right)$, the equation~\eqref{eqn_3} provided by Friis to estimate the stage-wise noise factors is not equivalent to our equation~\eqref{eqn_6} derived using the basic definition of a noise factor \cite{bib25,bib30,bib31,bib32}. From the above discussion, one of the major errors in Friis' stage-wise noise factor formula (especially for stages $x\geq2$) is that the `total noise powers at the input terminals of a cascade network's $x$-th stage' is erroneously considered as the `total noise power at the input terminals of the overall cascade network.' This error is not applicable for a single-stage network because the `total noise power at the input terminals of a cascade network's 1-st stage' is the same as the `total noise power at the input terminals of the overall cascade network.' Therefore, our new formula for the stage-wise noise factor is a correction to Friis' formula.

Although the stage-wise noise factor estimates differ for $x \geq 2$, the cascade network's total noise factor estimated using equations~\eqref{eqn_4} and~\eqref{eqn_9} remains the same for all cascade networks. This is because both equations are derived from the same base equation,~\eqref{eqn_2} or~\eqref{eqn_8}. This similarity is presented using a bar chart in Fig.~\ref{fig_2}c. However, since our formula for the stage-wise noise factor (equation~\eqref{eqn_6}) is a correction to Friis' formula (equation~\eqref{eqn_3}), our total noise factor formula for cascade networks in terms of their stage-wise noise factors (equation~\eqref{eqn_9}) will also be a correction to Friis' formula (equation~\eqref{eqn_4}).

\subsection{Comparison with Friis' noise factor formulas for cascade networks (neglecting externally added stage-wise noises)} 
\label{subsec3_2}

\begin{figure}[!t]
\centering
\includegraphics[width=5in]{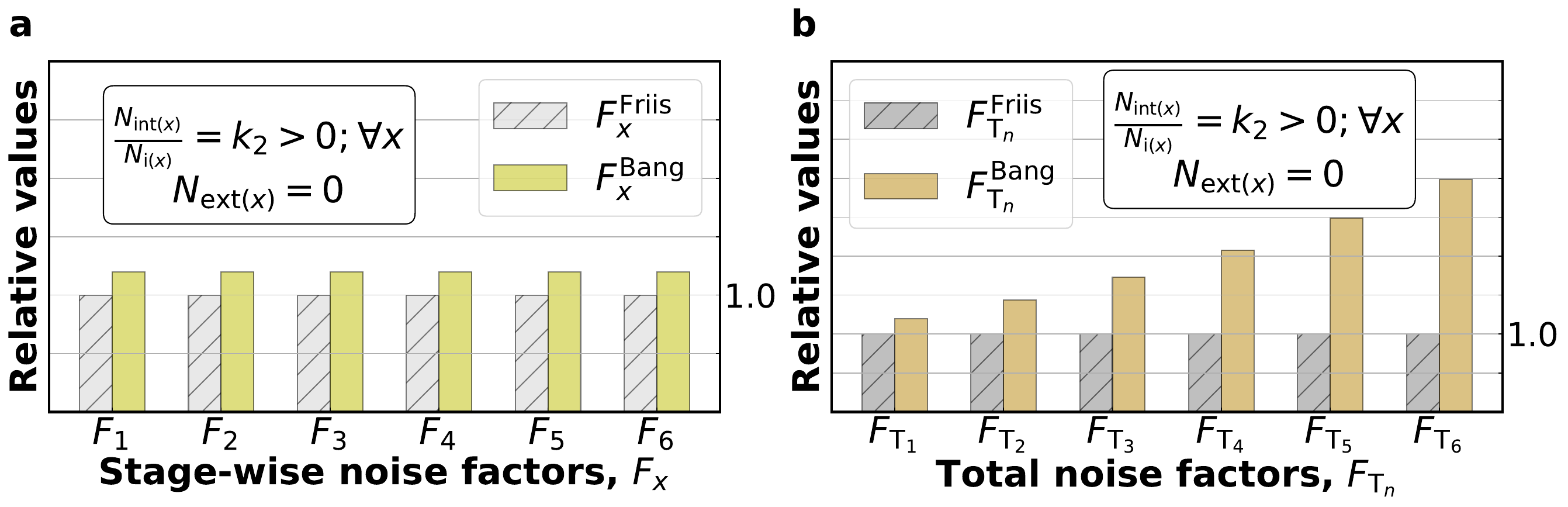}
\caption{A comparison of Friis' and our noise factors for up to the 6-th stage (neglecting externally added noises). Bar charts comparing Friis' and our (a) stage-wise noise factor estimates and (b) total noise factor estimates.}
\label{fig_3}
\end{figure}

Assuming that there are no externally added noises at the output of each stage in a cascade network, we obtain unity stage-wise Friis noise factors. Whereas, the stage-wise noise factors obtained using our formula (equation~\eqref{eqn_6}) will be greater than unity owing to the presence of internally generated stage-wise noises. However, in solid-state cascade devices such as staircase APDs, the stepwise internal noise generated due to the irregularities in the stepwise impact ionization is proportional to its stepwise input noise. Therefore, in cascade networks, we consider the ratio of stage-wise internal noise to stage-wise input noise as a constant. A comparison of stage-wise noise factor estimates is presented as a bar chart in Fig.~\ref{fig_3}a, considering equal stage-wise power gains and stage-wise internal noise to stage-wise input noise ratios. Further, a comparison of the corresponding relative total noise factor estimates is depicted in Fig.~\ref{fig_3}b. Here, we observe that Friis' total noise factors for an $n$-stage cascade network are unity. Whereas, our estimates are not just greater than unity but increase with the stage count. Therefore, as the number of stages increases, the difference in the values estimated using Friis' and our formulas increases drastically. The inequalities observed in the relative estimates of noise factors presented in Fig.~\ref{fig_3}a and Fig.~\ref{fig_3}b are because Friis' formulas do not include the stage-wise internally generated noises, which are a major source of noise in cascade devices such as multistep staircase APDs. This necessitates a correction to Friis' noise factor formulas.

\subsection{Correlation with staircase APD's excess noise factor expressions} 
\label{subsec3_3}

Staircase APDs are solid-state devices analogues to a cascade network with twofold stepwise gains via impact ionization when operated in their staircase operation regime. Therefore, the terms noise factor (total or stage-wise) used in this article for cascade networks are equivalent to the terms excess noise factor (total or stepwise) for staircase APDs. Table~\ref{tab1} correlates our newly derived noise factor expressions for a cascade network with the excess noise factor expressions for a staircase APD. Here, the results 1 and 2 are derived from the basic definitions for (excess) noise factors corresponding to two-port networks or semiconductor devices. Since the externally added noises are negligible in staircase APDs \cite{bib2,bib3,bib33}, here we ignore the term corresponding to $N_{\text{ext}(x)}$ for cascade networks.

\begin{table}[h]
\caption{Correlation between a cascade network and a staircase APD}\label{tab1}
\begin{tabular*}{\textwidth}{@{\extracolsep\fill}lcccc}
\toprule%
& \multicolumn{2}{@{}c@{}}{Cascade network\footnotemark[1]} & \multicolumn{2}{@{}c@{}}{Staircase APD\footnotemark[2]} \\ 
\cmidrule{2-3}\cmidrule{4-5}%
 & Terminology & Expression & Terminology & Expression \\
\midrule
Basic & Noise factor & \multirow{2}{*}{$F=\frac{\text{SNR}_\text{i}}{\text{SNR}_\text{o}}$} & Excess noise factor & \multirow{2}{*}{$F=\frac{\langle M^2 \rangle}{\langle M \rangle ^2}$}  \\
definition & (two-port network) &  & (semiconductor device) &   \\
\cmidrule{1-1}
\cmidrule{2-3}\cmidrule{4-5}
Derivation & \multicolumn{2}{@{}c@{}}{Formula-based} & \multicolumn{2}{@{}c@{}}{Statistical method of}  \\
method & \multicolumn{2}{@{}c@{}}{mathematical derivation} & \multicolumn{2}{@{}c@{}}{random process} \\
\midrule
\multirow{2}{*}{Result 1}  & Total & \multirow{2}{*}{$\prod_{x=1}^{n}F_x^{\text{Bang}}$} & Total & \multirow{2}{*}{$\prod_{x=1}^{n}F_x^{\text{S-APD}}$} \\
  & noise factor &  & excess noise factor &  \\
  &  &  &  & \multirow{2}{*}{$=\prod_{x=1}^{n}\left\{\frac{(1+3p_x)}{(1+p_x)^2}\right\}$}  \\
	&  &  &  &  \\
  &  &  &  & \multirow{2}{*}{$=\prod_{x=1}^{n}\left\{\frac{\langle M_x^2 \rangle }{\langle M_x \rangle ^2}\right\}$}  \\
	&  &  &  &  \\
\midrule
\multirow{2}{*}{Result 2}  & Stage-wise & \multirow{2}{*}{$F_x^\text{Bang} (\geq 1)$} & Stepwise &  \multirow{2}{*}{$F_x^\text{S-APD} (\geq 1)$} \\
  & noise factor &  & excess noise factor &  \\
  &  & \multirow{2}{*}{$= 1+\frac{N_{\text{int}(x)}}{N_{\text{i}(x)}G_x}$} &  & \multirow{2}{*}{$= 1+\frac{p_x(1-p_x)}{(1+p_x)^2}$}  \\
	&  &  &  &  \\
  &  & \multirow{2}{*}{$= 1+\delta_{\text{int}(x)}$} &  & \multirow{2}{*}{$= 1+\frac{\text{var}(M_x)}{\langle M_x \rangle ^2}$}  \\
	&  &  &  &  \\
\bottomrule
\end{tabular*}
\thefootnote{Note: $M$ is the gain of the semiconductor device; $p_x$ is the stepwise ionization probability of the staircase APD; $M_x$ is the stepwise gain of the staircase APD.} \\
\thefootnote{$^1$Refer subsection~\ref{subsec2_2}.} \\
\thefootnote{$^2$Reference~\cite{bib33}.}
\end{table}

Importantly, although the results in Table~\ref{tab1} are derived from two separate basic definitions for (excess) noise factors using two different derivation methods, result 1 shows that the total noise factor for a cascade network is equal to the product of its stage-wise noise factors, similar to a staircase APD. But, this contradicts Friis' total noise factor formula for a cascade network given by equation~\eqref{eqn_4}. Similarly, result 2 shows that the stage-wise (stepwise excess) noise factors are greater than or equal to unity, in both cases.

\section{Conclusion} 
\label{sec4}

We discussed the existing theory on noise factors proposed by Friis, which does not include the internally generated noise components. However, our re-derived formulas for stage-wise and total noise factors for a cascade network consider the stage-wise internal noises, which are a major source of noise in cascade devices such as a multistep staircase APD \cite{bib33}. We further highlighted the differences in the noise factor formulas derived by us and Friis and discussed the need for a correction in Friis' formulas. The key differences are noticed in the equations~\eqref{eqn_3} (Friis') and \eqref{eqn_6} (ours) for stage-wise noise factors and equations~\eqref{eqn_4} (Friis') and \eqref{eqn_9} (ours) for total noise factors. To validate our derived formulas, we correlated our new formulas with staircase APD's excess noise factor expressions. Remarkably, our new formula for a cascade network's total noise factor derived using a formula-based mathematical derivation is equal to the product of its stage-wise noise factors, similar to the staircase APD's expression derived using a statistical method of random process. Future work will focus on correlating our noise model with other cascade-amplifying devices like staircase APDs and expressing the cascade network's stage-wise internal noise powers in terms of device parameters such as an APD's stepwise ionization probabilities, stepwise gains etc. Our newly derived stage-wise and total noise factor formulas for multistage cascade networks can be used in modeling, predicting, and estimating noise in various cascade devices.

\section*{Acknowledgments}
A.E.B. would like to thank the Indian Institute of Technology Bombay for their support.

\section*{Appendices}
\appendix

\section{Derivation of Friis' total noise factor formula} 
\label{secA1}

According to Friis \cite{bib25}, the stage-wise noise factors of a cascade network is expressed as,  

\begin{equation}
\label{Appendix_A_eqn_1}
F_x^{\text{Friis}} = 1+\frac{N_{\text{ext}(x)}}{N_\text{i}G_x}
\end{equation}

Therefore, Friis' total noise factor formula for a cascade network in terms of its stage-wise noise factors (neglecting the stage-wise internal noise powers ($N_{\text{int}(x)}$)) is given by, 

\begin{equation}
\label{Appendix_A_eqn_2}
\begin{aligned}
F_{\text{T}_n}^{\text{Friis}} &= \frac{\text{SNR}_\text{i}}{\text{SNR}_\text{o}} = \frac{\left({S_\text{i}}/{N_\text{i}}\right)}{\left({S_\text{o}}/{N_\text{o}}\right)} \\
&= \underbrace{1+\frac{N_{\text{ext}(1)}}{N_\text{i}G_1}}_{\text{$=F_1^{\text{Friis}}$}}+\underbrace{\frac{N_{\text{ext}(2)}}{N_\text{i}G_1G_2}}_{\text{$=\left(F_2^{\text{Friis}}-1\middle)\middle/\middle(G_1\right)$}} \\ 
&~+...+\underbrace{\frac{N_{\text{ext}(n)}}{N_\text{i}G_1G_2...G_n}}_{\text{$=\left(F_n^{\text{Friis}}-1\middle)\middle/\middle(\prod_{x=1}^{(n-1)}G_x\right)$}} \\ 
&= F_1^{\text{Friis}}+\sum_{x=2}^{n}\left(\frac{F_x^{\text{Friis}}-1}{\prod_{y=1}^{(x-1)}G_y}\right)
\end{aligned}
\end{equation}

\section{Derivation of our new total noise factor formula} 
\label{secB1}

Our new expression for a cascade network's stage-wise noise factors is given by, 

\begin{equation}
\label{Appendix_B_eqn_1}
\begin{aligned}
F_x^{\text{Bang}} = 1+\frac{\left(N_{\text{int}(x)}+N_{\text{ext}(x)}\right)}{N_{\text{i}(x)}G_x} 
\end{aligned}
\end{equation}

Rewriting the stage-wise noise factor at the $x$-th stage of a cascade network in terms of its previous stages' stage-wise noise factors, 

\begin{equation}
\label{Appendix_B_eqn_2}
F_x^{\text{Bang}} = 1+\frac{\left(N_{\text{int}(x)}+N_{\text{ext}(x)}\right)}{N_\text{i}\prod_{j=1}^{x}G_j\prod_{k=1}^{(x-1)}F_k^{\text{Bang}}}; \forall x\geq2 
\end{equation}

Therefore, rearranging equation~\eqref{Appendix_B_eqn_2} and substituting it in equation~\eqref{Appendix_B_eqn_3}, the new total noise factor formula for a cascade network in terms of its stage-wise noise factors (including the stage-wise internal noise powers ($N_{\text{int}(x)}$)) is given by, 

\begin{equation}
\label{Appendix_B_eqn_3}
\begin{aligned}
F_{\text{T}_n}^{\text{Bang}} &= \frac{\text{SNR}_\text{i}}{\text{SNR}_\text{o}} = \frac{\left({S_\text{i}}/{N_\text{i}}\right)}{\left({S_\text{o}}/{N_\text{o}}\right)} \\ 
=& \underbrace{1+\frac{\left(N_{\text{int}(1)}+N_{\text{ext}(1)}\right)}{N_\text{i}G_1}}_{\text{$=F_1^{\text{Bang}}$}}+\underbrace{\frac{\left(N_{\text{int}(2)}+N_{\text{ext}(2)}\right)}{N_\text{i}G_1G_2}}_{\text{$=\left(F_2^{\text{Bang}}-1\right)F_1^{\text{Bang}}$}}\\
&+...+\underbrace{\frac{\left(N_{\text{int}(n)}+N_{\text{ext}(n)}\right)}{N_\text{i}G_1G_2...G_n}}_{\text{$=\left(F_n^{\text{Bang}}-1\right)\prod_{x=1}^{n-1}F_x^{\text{Bang}}$}} \\ 
&= \prod_{x=1}^{n}F_x^{\text{Bang}}\neq\Bigl(F_{\text{T}_n}^{\text{Friis}}\Bigr) 
\end{aligned}
\end{equation}

\section{Illustration to prove the inequality in Friis' and ours stage-wise noise factor formulas, for higher stages (neglecting internally generated stage-wise noises)} 
\label{secC1}

The noise factor a single-stage network or the stage-wise noise factor at the 1-st stage of a cascade network is given by, 

\begin{equation}
\label{Appendix_C_eqn_1}
\begin{aligned}
F_1^{\text{Bang}} &= \frac{\text{SNR}_{\text{i}(1)}}{\text{SNR}_{\text{o}(1)}} = \frac{\text{SNR}_{\text{i}}}{\text{SNR}_{\text{o}(1)}} \\
&=\frac{N_{\text{o}(1)}}{N_{\text{i}(1)}G_1} = 1 + \frac{N_{\text{ext}(1)}}{N_\text{i}G_1} \\ 
&=\left(F_1^{\text{Friis}}=1+\frac{N_{\text{ext}(1)}}{N_\text{i}G_1}\right) 
\end{aligned}
\end{equation}

However, considering the noise factor at the 2-nd stage of a network,

\begin{equation}
\label{Appendix_C_eqn_2}
\begin{aligned}
F_2^{\text{Bang}} &= \frac{\text{SNR}_{\text{i}(2)}}{\text{SNR}_{\text{o}(2)}} = \frac{\text{SNR}_{\text{o}(1)}}{\text{SNR}_{\text{o}(2)}} \\
&=\frac{N_{\text{o}(2)}}{N_{\text{i}(2)}G_2} = 1 + \frac{N_{\text{ext}(2)}}{\left(N_\text{i}G_1+N_{\text{ext}(1)}\right)G_2} \\ 
&\neq\left(F_2^{\text{Friis}}=1+\frac{N_{\text{ext}(2)}}{N_\text{i}G_2}\right) 
\end{aligned}
\end{equation}

From equation~\eqref{Appendix_C_eqn_2}, it is observed that $N_\text{i} \neq N_{\text{i}(2)} \left(=N_\text{i}G_1+N_{\text{ext}(1)}\right)$. This inequality shown in equation~\eqref{Appendix_C_eqn_2} applies to all higher stages.

\end{document}